\documentclass[%
 reprint,
superscriptaddress,
 amsmath,amssymb,
 aps,
prb,
]{revtex4-2}

\usepackage{graphicx}
\usepackage{dcolumn}
\usepackage{bm}
\usepackage{siunitx}
\usepackage{hyperref}
\usepackage{placeins}
\usepackage[T1]{fontenc}
\usepackage{etoolbox}
\usepackage{xcolor}
\usepackage{mathptmx}
\usepackage{nameref}
\usepackage{booktabs}
\usepackage{chemformula}
\usepackage{ulem}

\newcommand{\rev}[1]{\textcolor{black}{#1}}

\DeclareSIUnit{\rad}{rad}

\hypersetup{
        colorlinks   = true,
        citecolor    = blue,
        linkcolor    = blue}

\begin{document}

\title{Anomalous Hall effect and magnetoresistance in micro-ribbons\\ of the magnetic Weyl semimetal \rev{candidate} \ch{PrRhC2}}


\author{Mickey Martini}
  \email{m.martini@ifw-dresden.de}
 \affiliation{Leibniz Institute for Solid State and Materials Science Dresden (IFW Dresden), 01069 Dresden, Germany}
 \affiliation{Institute of Applied Physics, Technische Universität Dresden, 01062 Dresden, Germany}

   \author{Helena Reichlova}
 \affiliation{Institute of Physics of the Czech Academy of Sciences,
Na Slovance 1999/2, 18221 Prague, Czechia}
 \affiliation{Institut für Festkörper- und Materialphysik (IFMP), Technische Universität Dresden, 01069 Dresden, Germany}

 \author{Laura T. Corredor}
\affiliation{Leibniz Institute for Solid State and Materials Science Dresden (IFW Dresden), 01069 Dresden, Germany}

   \author{Dominik Kriegner}
 \affiliation{Institute of Physics of the Czech Academy of Sciences,
Na Slovance 1999/2, 18221 Prague, Czechia}
 
 \author{Yejin Lee}
 \affiliation{Leibniz Institute for Solid State and Materials Science Dresden (IFW Dresden), 01069 Dresden, Germany}
 \affiliation{Institute of Applied Physics, Technische Universität Dresden, 01062 Dresden, Germany}
 
  \author{Luca Tomarchio}
  \affiliation{Leibniz Institute for Solid State and Materials Science Dresden (IFW Dresden), 01069 Dresden, Germany}
 \affiliation{Department of Physics, Sapienza University, 00185 Rome, Italy}
 \affiliation{INFN Section of Rome, 00185 Rome, Italy}

 \author{Kornelius Nielsch}
 \affiliation{Leibniz Institute for Solid State and Materials Science Dresden (IFW Dresden), 01069 Dresden, Germany}
 \affiliation{Institute of Applied Physics, Technische Universität Dresden, 01062 Dresden, Germany}
 \affiliation{Institute of Materials Science, Technische Universität Dresden, 01062 Dresden, Germany}

 \author{Ali G. Moghaddam}
 \affiliation{Leibniz Institute for Solid State and Materials Science Dresden (IFW Dresden), 01069 Dresden, Germany}
\affiliation{Department of Physics, Institute for Advanced Studies in Basic Sciences (IASBS), Zanjan 45137-66731, Iran}
\affiliation{Computational Physics Laboratory, Physics Unit, Faculty of Engineering and Natural Sciences, Tampere University, FI-33014 Tampere, Finland}


\author{Jeroen van den Brink}
 \affiliation{Leibniz Institute for Solid State and Materials Science Dresden (IFW Dresden), 01069 Dresden, Germany}
\affiliation{Institute for Theoretical Physics and Würzburg-Dresden Cluster of Excellence ct.qmat, Technische Universität Dresden, 01069 Dresden, Germany}

\author{Bernd Büchner}
 \affiliation{Leibniz Institute for Solid State and Materials Science Dresden (IFW Dresden), 01069 Dresden, Germany}
 \affiliation{Institut für Festkörper- und Materialphysik (IFMP), Technische Universität Dresden, 01069 Dresden, Germany}
 
\author{Sabine Wurmehl}
 \affiliation{Leibniz Institute for Solid State and Materials Science Dresden (IFW Dresden), 01069 Dresden, Germany}

 \author{Vitaliy Romaka}
 \affiliation{Leibniz Institute for Solid State and Materials Science Dresden (IFW Dresden), 01069 Dresden, Germany}
 
  \author{Andy Thomas}
     \email{a.thomas@ifw-dresden.de}
 \affiliation{Leibniz Institute for Solid State and Materials Science Dresden (IFW Dresden), 01069 Dresden, Germany}
 \affiliation{Institut für Festkörper- und Materialphysik (IFMP), Technische Universität Dresden, 01069 Dresden, Germany}
 
\received{\today}
             

\begin{abstract}
\ch{PrRhC2} belongs to the rare-earth carbides family whose properties are of special interest among topological semimetals due to the simultaneous breaking of both inversion and time-reversal symmetry. The concomitant absence of both symmetries grants the possibility to tune the Weyl nodes chirality and to enhance topological effects like the chiral anomaly. In this work, we report on the synthesis and compare the magnetotransport measurements of a poly- and single crystalline \ch{PrRhC2} sample. Using a remarkable and sophisticated technique, the \ch{PrRhC2} single crystal is prepared via focused ion beam cutting from the polycrystalline material. Our magnetometric and specific heat analyses reveal a non-collinear antiferromagnetic state below $\SI{20}{\kelvin}$, as well as short-range magnetic
correlations and/or magnetic fluctuations well above the onset of the magnetic transition. The transport measurements on the \ch{PrRhC2} single crystal display an electrical resistivity peak at $\SI{3}{\kelvin}$ and an anomalous Hall effect below $\SI{6}{\kelvin}$ indicative of a net magnetization component in the ordered state. Furthermore, we study the angular variation of magnetoresistivities as a function of the angle between the in-plane magnetic field and the injected electrical current. We find that both the transverse and the longitudinal resistivities exhibit fourfold angular dependencies due to higher-order terms in the resistivity tensor, consistent with the orthorhombic crystal symmetry of \ch{PrRhC2}. Our experimental results may be interpreted as features of topological Weyl semimetallic behavior in the magnetotransport properties.

\end{abstract}

\pacs{Valid PACS appear here}

\maketitle

\section{\label{sec:level1}Introduction}

 \begin{figure*}[t!]
    \centering
    \includegraphics[width=.88\textwidth]{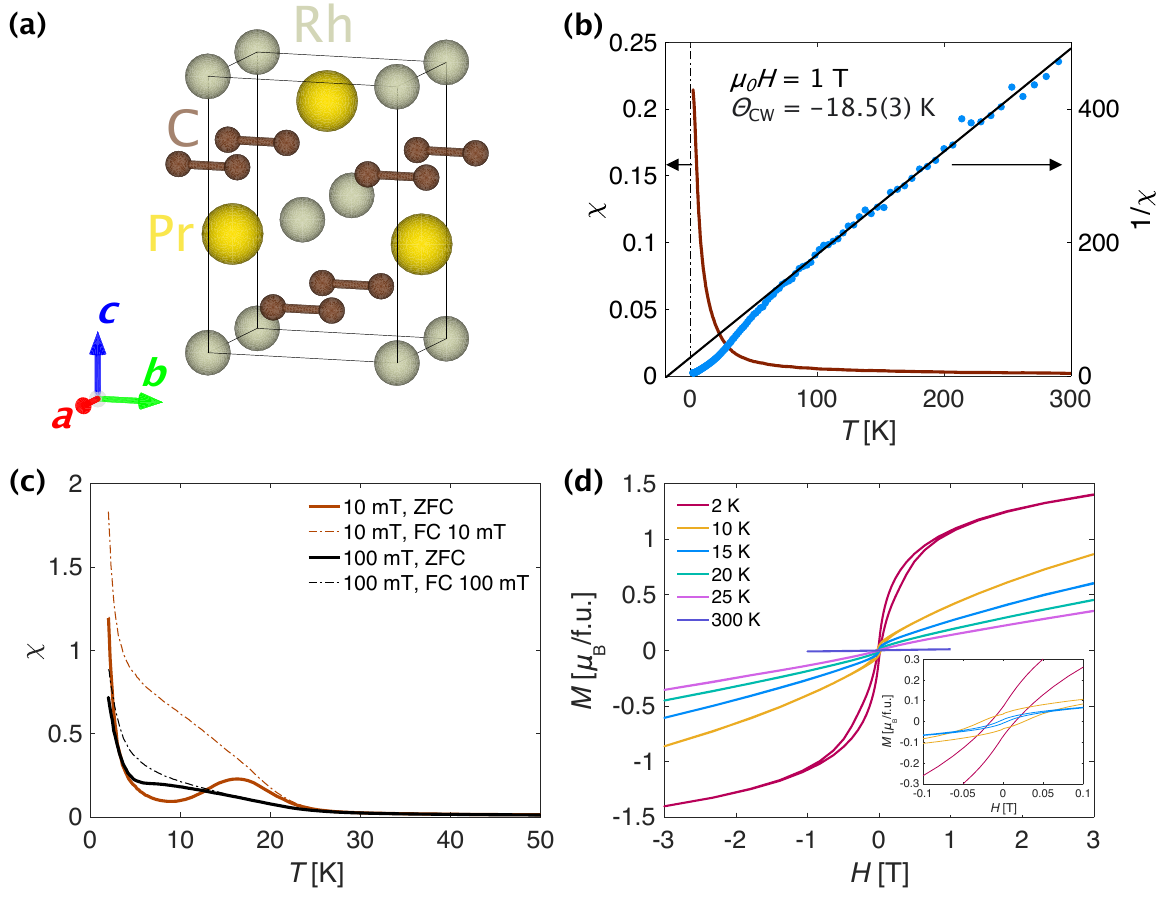}
    \caption{(a) Crystal structure of \ch{PrRhC2} (orthorhombic symmetry, spacegroup $Amm2$) including the expected Carbon-dimers. The unit cell is indicated by black solid lines. (b) Magnetic susceptibility (SI units) of \ch{PrRhC2} polycrystal measured while increasing the temperature in presence of external magnetic field of $\SI{1}{\tesla}$ (brown curve) and reciprocal of the susceptibility versus temperature (blue dots) and a Curie-Weiss fit (CW) to the data (solid black line). This linear fit yields the CW temperature $\Theta_\mathrm{CW}$ to be about -18.5~K. (c) Magnetic susceptibility versus temperatures recorded in different magnetic fields following a zero-field-cooled (ZFC) and field-cooled (FC) protocol. (d) Magnetization curves as a function of magnetic field at different temperatures. Inset: Magnetization at low field and low temperatures ($\SI{2}{\kelvin}$, $\SI{10}{\kelvin}$, $\SI{15}{\kelvin}$). }
    \label{fig:SQUID}
\end{figure*}

The realization of Weyl fermions in condensed matter systems is of great interest to prove novel topological theories while also having promising applications in the field of quantum computing, spintronics and electro-optical devices \cite{wan2011topological,burkov2011weyl,wang2012dirac,liu2014weyl, weng2015weyl,burkov2016topological, yan2017topological, armitage2018weyl,morali2019fermi, sie2019ultrafast}. Weyl nodes form a pair by breaking a symmetry such as inversion and/or time-reversal. Although the majority of the, so far, discovered Weyl semimetals (WSMs) is of the parity-broken type, many experimental attempts have been performed to find magnetic WSMs that, contrary to the non-magnetic counterparts, combine magnetic order and a topologically non-trivial band character, giving rise to exotic phenomena such as the intrinsic anomalous Hall effect and axion electrodynamics \cite{kuroda2017evidence, kim2018large, liu2018giant,liu2019magnetic, shen2019anisotropies, okamura2020giant, su2020magnetic, nie2022magnetic}. Among the proposed magnetic WSMs, rare-earth carbides \cite{ray2022tunable, sadhukhan2023effect} break both inversion and time-reversal symmetry due to their non-centrosymmetric point group and magnetic order, respectively \cite{xu2017discovery,chang2018magnetic,yang2020transition,yang2021noncollinear,zhang2022weyl}. 

They represent a promising material class to probe the "chiral anomaly" effect, considered a key fingerprint of topology. In presence of parallel magnetic and electric fields, the electrons flow between Weyl nodes of opposite chirality thereby inducing a negative quadratic longitudinal magnetoresistance (MR) \cite{nielsen1983adler,zyuzin2012topological, son2013chiral, huang2015observation}. This chiral imbalance between the Weyl nodes is permitted in materials where both inversion and time-reversal symmetries are absent causing the net chirality picked up by experimental probes to vanish and, hence, the  "chiral anomaly" effect cannot be seen. In turn, observation of a negative longitudinal MR, whose magnitude strongly depends on the angle between magnetic and electric fields, is considered a key observation indicative of Weyl physics via a topologically protected chiral charge \cite{jia2016weyl}. As an alternative description, a negative MR can also be the result of extrinsic mechanisms, such as weak electron localization \cite{zhang2017tunable}, ionic impurity scattering processes \cite{goswami2015axial} or conductivity fluctuations \cite{schumann2017negative}, making the identification of the anomaly highly non-trivial. In semimetals with high mobility, the evidence of a chiral anomaly can also be strongly hindered by the "current jetting" effect that leads to a strong apparent negative longitudinal MR \cite{dos2016search, arnold2016negative}.

In the class of rare-earth carbide materials, it has been predicted that an appropriate tilt of the magnetization can stabilize an odd number of Weyl nodes near the Fermi surface, giving a nonzero net chirality that can be easily experimentally probed through the chiral anomaly effect \cite{ray2022tunable, sadhukhan2023effect}. The class of rare-earth/transition-metal carbides, that, according to the theory, combine topology and magnetism, includes \ch{GdCoC2}, \ch{GdNiC2}, \ch{NdRhC2} and \ch{PrRhC2} \cite{ray2022tunable}. Magnetization measurements of single crystals reveal that \ch{GdCoC2} is antiferromagnetic (AFM) below Néel temperature $T_N = \SI{15.6}{\kelvin}$ \cite{matsuo1996antiferromagnetism} and shows an order-order transition at $T= \SI{14.0}{\kelvin}$. Reference \cite{meng2016magnetic} descrives the same material as a ferromagnet with Curie temperature $T_C\simeq \SI{15}{\kelvin}$. On the other hand, \ch{GdNiC2} single crystals, as most of the Ni-based members in this family \cite{onodera1998magnetic}, were proven to reach an antiferromagnetic state below $T_N = \SI{20}{\kelvin}$ \cite{matsuo1996antiferromagnetism}. Experimental reports on polycrystals of \ch{NdRhC2} and \ch{PrRhC2} are until now limited to their high-temperature susceptibility \cite{hoffmann1989structural} and single crystals of \ch{NdRhC2} and \ch{PrRhC2} have not been grown, so far. The high-temperature susceptibility measurements suggest that these two compounds possess an AFM ground state with critical temperature of $\sim\SI{13}{\kelvin}$ and close to \SI{0}{\kelvin}, respectively.  Density functional calculations based on generalized gradient approximation have predicted, however, a FM ground state for \ch{NdRhC2}, whereas \ch{PrRhC2}, is theoretically found to exhibit either an AFM or FM ground state, depending on the simulation approximations for the two unpaired $4f$-electrons \cite{ray2022tunable}. Therefore, the growth of a \ch{PrRhC2} single crystal is highly desirable to experimentally verify its magnetic structure.\newline \newline
Here, we circumvent this experimental challenge by cutting single crystal micro-ribbons from a multi-grain sample by focused ion beam. In addition to analyze the magnetic order of the polycrystalline \ch{PrRhC2} material through magnetometric measurements, we are therefore able to investigate the magnetotransport in the single crystal. Since for the PrRhC\textsubscript{2} crystal structure only ferromagnetic and non-collinear antiferromagnetic order \cite{chen2014anomalous,nakatsuji2015large} produces anomalous Hall effect, this lets us conclude on the \rev{type of magnetic order}.

 \begin{figure*}[]
    \centering
    \includegraphics[width=.45\textwidth]{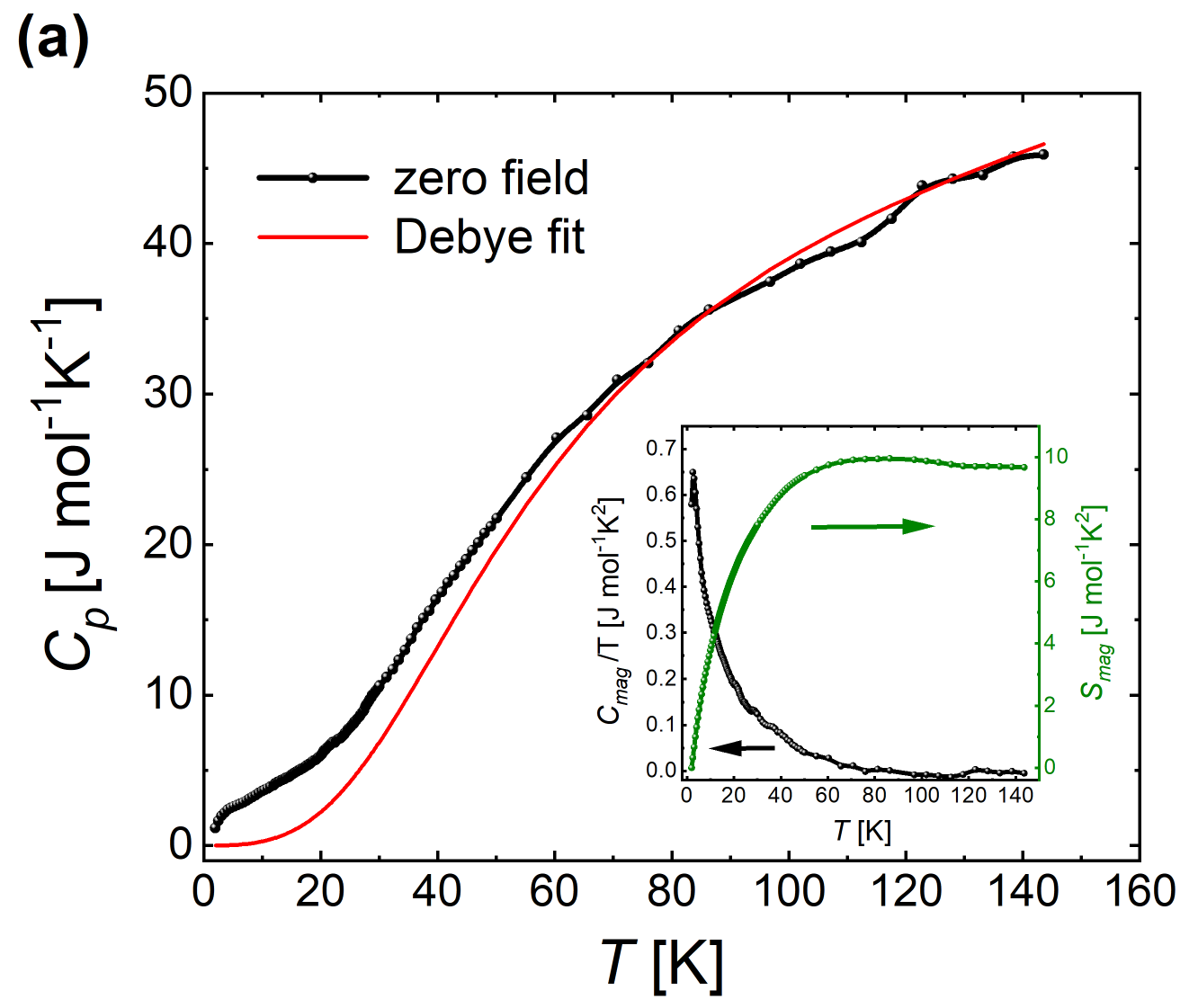}\quad
     \includegraphics[width=.45\textwidth]{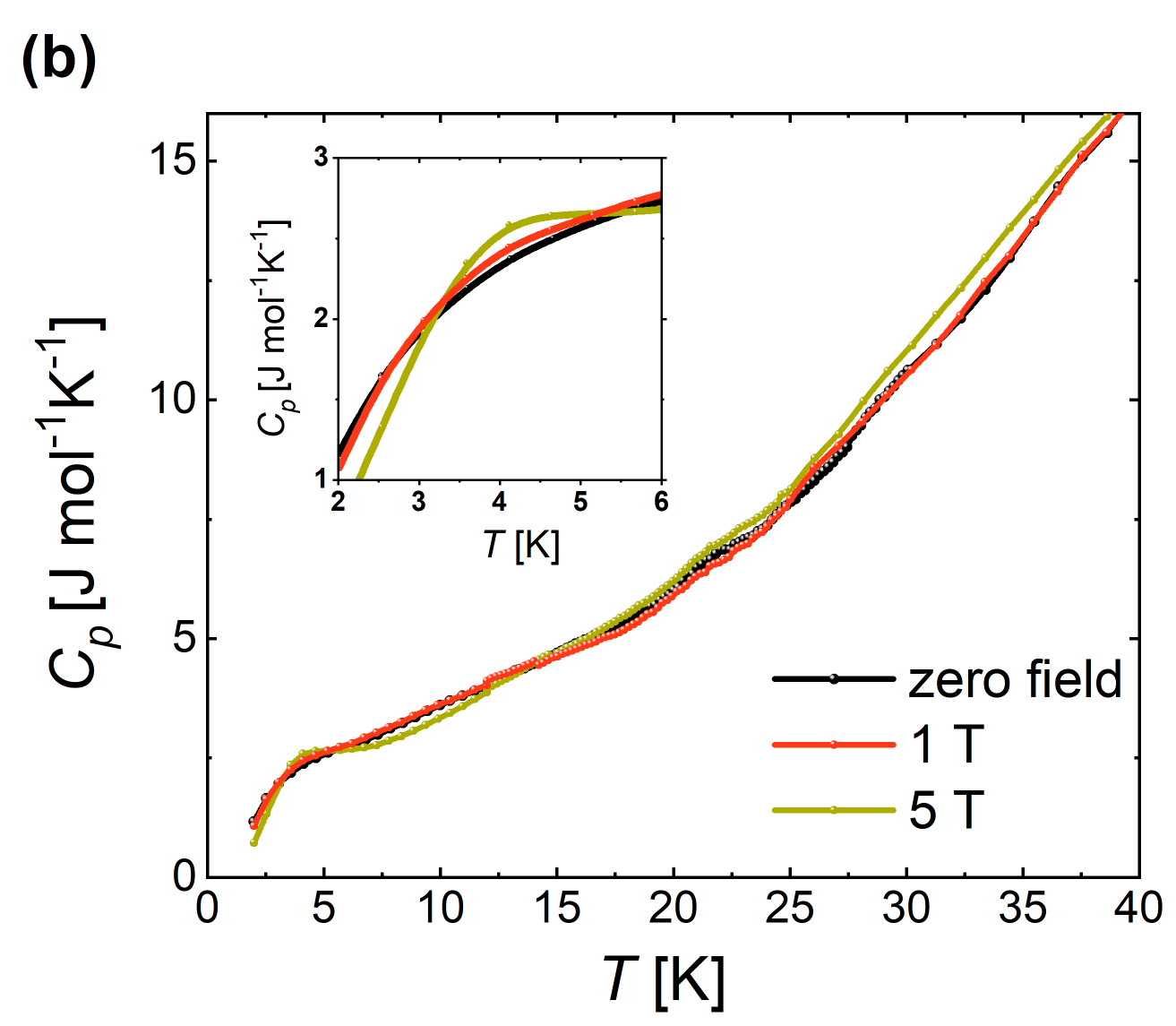}
    \caption{(a) Zero-field specific heat of \ch{PrRhC2}. The red line represents the lattice specific heat obtained using a Debye model; for details see text. Inset: zero-field magnetic specific heat plotted as $C_{mag}/T$ vs $T$ (left scale) together with the magnetic entropy (right scale). (b) Temperature dependence of the specific heat for applied magnetic fields of $\SI{1}{\tesla}$ and $\SI{5}{\tesla}$. The inset shows a zoom of the low temperature region.}
    \label{fig:HeatCapacity}
\end{figure*}

{\section{Preparation and magnetic characterization of the polycrystal}\label{Section:preparation_poly}

Polycrystalline samples are prepared by arc melting slugs/pieces of > 99.9 mass\% metals-based pure elements on a water-cooled copper hearth under purified argon. The metal blends are melted three times for homogenization keeping the evaporation losses below 1 wt.\%. Samples are then individually wrapped in protective Ta-foil, sealed in evacuated silica ampules, and annealed at $\SI{1000}{\celsius}$ for 1 week followed by quenching the ampules in cold water. The crystal structure of polycrystalline samples is determined by X-ray powder diffraction analysis (STOE STADI in transmission geometry, CoK$\alpha$1 radiation, equipped with a germanium monochromator and a DECTRIS MYTHEN 1K detector). The lattice parameters are obtained from a structural model derived by Rietveld analysis yielding: $a$ = 0.36876(1), $b$ = 0.47103(1), $c$ = 0.66124(1)\,nm, $R_{p}$ = 0.0281, $R_{wp} = 0.0371$, $R_I$ = 0.0563, $\chi^2$ = 1.28. \rev{More details can be found in Tab.~S1 in the Supplementary Information.}

\begin{figure*}[t!]
\centering
    \includegraphics[width=1\textwidth]{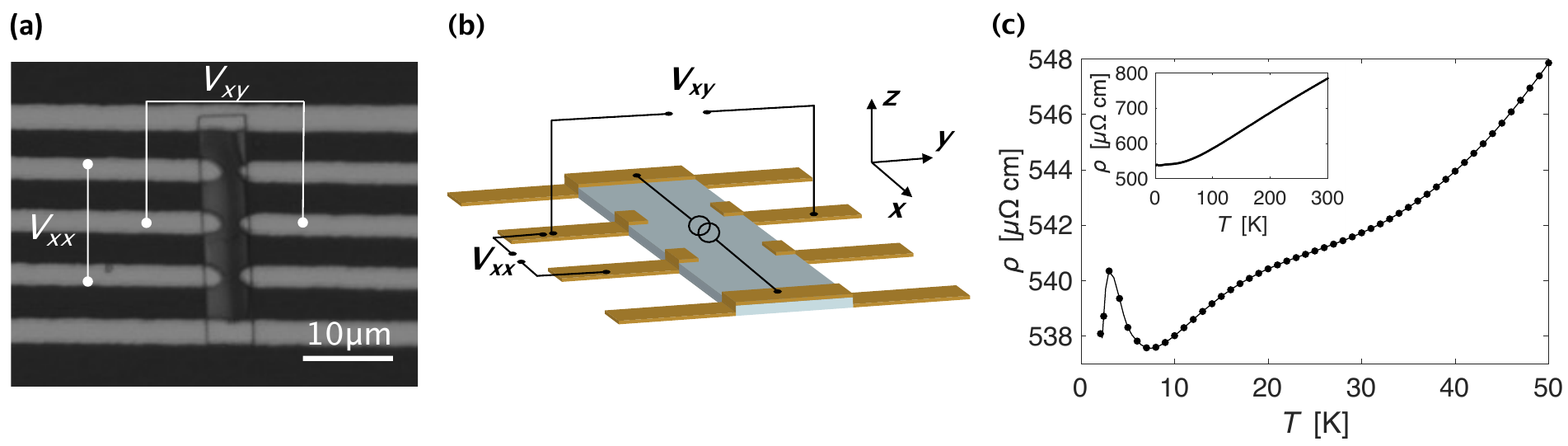}
\caption{(a) Closeup microscopy image of the device. The lamella is obtained with a FIB cut and electrical contacts are defined by standard optical lithography. (b) Schematics of the device with coordinate system. (c) Electrical resistivity measured in a four points geometry at low temperatures (main panel) and in the full range of temperatures (inset).}
    \label{fig:device}
\end{figure*}

The magnetic structure of the polycrystal is investigated by magnetometry measurements in a superconducting quantum interference device (SQUID) down to $\SI{2}{\kelvin}$. The magnetic susceptibility $\chi$ of \ch{PrRhC2}\, [Fig.~\hyperref[fig:SQUID]{1(b)}], measured at a $\SI{1}{\tesla}$-field from $2$ to $\SI{300}{\kelvin}$, displays a steep increase below $\SI{10}{\kelvin}$, indicating a net magnetization at low temperature. The inverse susceptibility above $\SI{50}{\kelvin}$ is well fitted by the Curie-Weiss (CW) law $1/\chi = (T-\Theta_{CW})/C$, where $C$ is related to the effective magnetic moment per unit cell, and $\Theta_{CW}$ is the CW temperature, connected to to the sum of all magnetic interactions in the system. The obtained effective magnetic moment $\mu_{eff} = \SI{3.7(2)}\mu_{B}$ is in agreement with the expected for Pr$^{3+}$, $\SI{3.58}\mu_{B}$ \cite{Blundell}. The CW temperature is negative ($\Theta_{CW} = -\SI{18.5(3)}{\kelvin}$), indicating AFM interactions. On the other hand, the magnetic susceptibility [Fig.~\hyperref[fig:SQUID]{1(c)}], measured from low to high temperature in a $\SI{10}{\milli\tesla}$ and $\SI{100}{\milli\tesla}$-fields display a pronounced ZFC/FC splitting starting around $\SI{24}{\kelvin}$. In the FC curve, a ferromagnetic-like transition is observed at T$\approx\SI{22}{\kelvin}$ (determined using the second derivative method, i.e., the transition temperature is defined as the inflection point of the curve). For the zero-field-cooled (ZFC) magnetization, a kink between $\SI{10}{\kelvin}$ and $\SI{20}{\kelvin}$ possibly related to a rearrangement of the magnetic structure, is rapidly suppressed by applying an external magnetic field. Figure \hyperref[fig:SQUID]{1(d)} depicts the magnetization as a function of external magnetic fields, at different base temperatures, which indicates the ferromagnetic-like nature of the compound, exhibiting clear hysteresis loops with coercive fields smaller than $\SI{30}{\milli\tesla}$ below $\SI{15}{\kelvin}$. Additionally, by increasing the applied magnetic field, the magnetization at the lowest temperature measured, $\SI{2}{\kelvin}$, approaches the value of $\SI{1.4}\mu_{B}$, which is below the expected $\SI{2}\mu_{B}$ for the $4f^2$ ($J=4$) state, assuming that Rh atoms have no magnetic moment. Altogether, our observations point to a non-collinear, canted AFM ground state with finite net magnetization which, nonetheless, is not fully saturated even \rev{up to $\SI{7}{\tesla}$ (Fig.~S2 in Supplementary Information). This finding makes the presence of magnetic domains unlikely. In ferromagnetic materials with well-organized domains, the magnetization reaches a saturation point where it no longer increases with an increasing magnetic field, corresponding to a single domain phase. Further Kerr microscopy and neutron diffraction experiments are necessary to investigate the saturation field and to understand the microscopic nature of the magnetic ordering.} This is, however, out of the scope of this work. It is worth noticing that we observe the same magnetic properties in two additional polycrystalline samples, whose surfaces are treated with CH\textsubscript{3} and HCl, respectively (not shown). 

The magnetic behavior of \ch{PrRhC2} is further analyzed by specific heat measurements as a function of temperature and applied magnetic field. The measurements are performed on four single crystals using a heat-pulse relaxation method in a Physical Property Measurement System (PPMS) from Quantum Design. In order to obtain the intrinsic specific heat, the temperature- and field-dependent addenda are thoroughly subtracted from the measured data. The temperature dependent specific heat is shown in Fig.\,\hyperref[fig:HeatCapacity]{2}. As can be observed in Fig.\,\hyperref[fig:HeatCapacity]{2(a)}, $C_p(T)$ at zero field does not show a clear anomaly associated with the magnetic transition observed in magnetic susceptibility around $\SI{22}{\kelvin}$: instead, a subtle bump appears, as well as a broad feature detected below $\SI{5}{\kelvin}$. In order to better understand this behavior and disentangle the magnetic contribution to the heat capacity, the lattice contribution is estimated using the following Debye model:

\begin{equation}\label{Debye}
  C_\textrm{ph}= 9R\sum_{i=1}^{2}n_i\left(\frac{T}{\Theta_{\textrm{D}i}}\right)^3\int_{0}^{\Theta_{\textrm{D}i}/T}\frac{x^4e^x}{(e^x-1)^2}dx ,
\end{equation}

where $R$ is the universal gas constant, the index $i$ sums over two different Debye temperatures $\Theta_{\textrm{D}i}$, each one representing a part of the total unit cell formed by $n$ = 4 atoms ($n_1$ = 2 and $n_2$ = 2). The values $\Theta_{\textrm{D}1} = \SI{240}{\kelvin} $ and $\Theta_{\textrm{D}2} = \SI{1540}{\kelvin}$ are determined as the best parameters. The total estimated lattice contribution is shown by the red line in Fig.\,\hyperref[fig:HeatCapacity]{2(a)}. By subtracting the lattice contribution thus calculated from the experimental data, the magnetic contribution to the specific heat is obtained, which is plotted as $C_{mag}/T$ as a function of temperature in the inset of  Fig.\,\hyperref[fig:HeatCapacity]{2(a)} (left axis). The magnetic entropy $S_{mag}$ is evaluated by integrating $C_{mag}/T$ [Fig.\,\hyperref[fig:HeatCapacity]{2(a)}, right axis]. An entropy release is observed up to $T$ $\sim$ 70~K, followed by a plateau with $S_{mag} \approx$ \qty[mode = text]{9.6}{J.mol^{-1}.K^{-1}}. Despite the limits of our phononic background, this value is close to the expected for a $S = 1$ ground state, (R $\log{(2S + 1)})$ = \qty[mode = text]{9.1}{J.mol^{-1}.K^{-1}}, probably overestimated by the non-perfect estimation of the lattice contribution. Most significant is the fact that magnetic fluctuations and/or short-range correlations seem to be present well above T$\approx\SI{22}{\kelvin}$, and that the entropy related to the magnetic order is gradually released over a wide temperature range. Last but not least, the temperature dependence of the specific heat under applied magnetic field is shown in Fig.\,\hyperref[fig:HeatCapacity]{2(b)}. While the subtle bump around $\SI{22}{\kelvin}$ does not change noticeably, the broad feature below $\SI{5}{\kelvin}$ slightly shifts to higher temperature, consistent with a robust ferromagnetic-like character of the canted AFM moments ordering at low temperature.

\section{Preparation and magneto-transport of the single crystal}\label{Section:preparation_single}

For magnetotransport studies, a single crystal of $\SI{28}{\micro\meter}\times\SI{6}{\micro\meter}\times\SI{300}{\nano\meter}$ size is mechanically extracted from the as-cast alloy and the relative crystal structure is determined using single-crystal X-ray diffraction, that confirms the orthorhombic space group $Amm2$ phase with lattice parameters consistent with the ones from the polycrystal. The microstructure and chemical composition of single and polyctystalline samples are analyzed by SEM on a ZEISS - EVO MA 15 equipped with an energy-dispersive X-ray (EDX) detector operated at $\SI{30}{\kilo\volt}$. Carbon is rather light element, so any quantification with EDX is not reliable. Therefore, only the Pr and Rh compositions are determined at several positions of each sample (Pr 50.2(4), Rh 49.8(4) at.\%) confirming their 1:1 ratio in agreement with the \ch{PrRhC2} composition. \rev{For cutting the micro-ribbon from the bulk material, we utilize a focused ion beam (FIB) of Ga\textsubscript{2+} ions. A typical position to yield a single crystal is indicated in Fig.~S1 in the Supplementary Information. After FIB preparation, the ribbons are lifted out of the crystal and transferred to the substrate for further device preparation. Details on the FIB preparation procedure can be found in  Ref.~\cite{geishendorf2019magnetoresistance}.}

The electrical contacts are defined by optical lithography and lift-off process with $\SI{300}{\nano\meter}$ of sputtered Au on thin Cr adhesive layer [Fig.~\hyperref[fig:device]{3(a)}]. Figure \hyperref[fig:device]{3(b)} shows the device layout with the electrical connections and the coordinate system. Figure \hyperref[fig:device]{3(c)} reports the electrical resistivity $\rho$ as a function of temperature measured using a four-point geometry. This dataset shows a metallic behavior with a hump around $\SI{20}{\kelvin}$, followed by a peak at $\SI{3}{\kelvin}$. These features go along with those observed in the magnetic susceptibility and specific heat measurements. The local increase of the electrical resistivity  at low temperature could be attributed to the enhanced carriers scattering by spin fluctuation due to rearrangement of the magnetic structure \cite{kataoka2001resistivity}.  

\begin{figure}[t!]
    \centering
    \includegraphics[width=.42\textwidth]{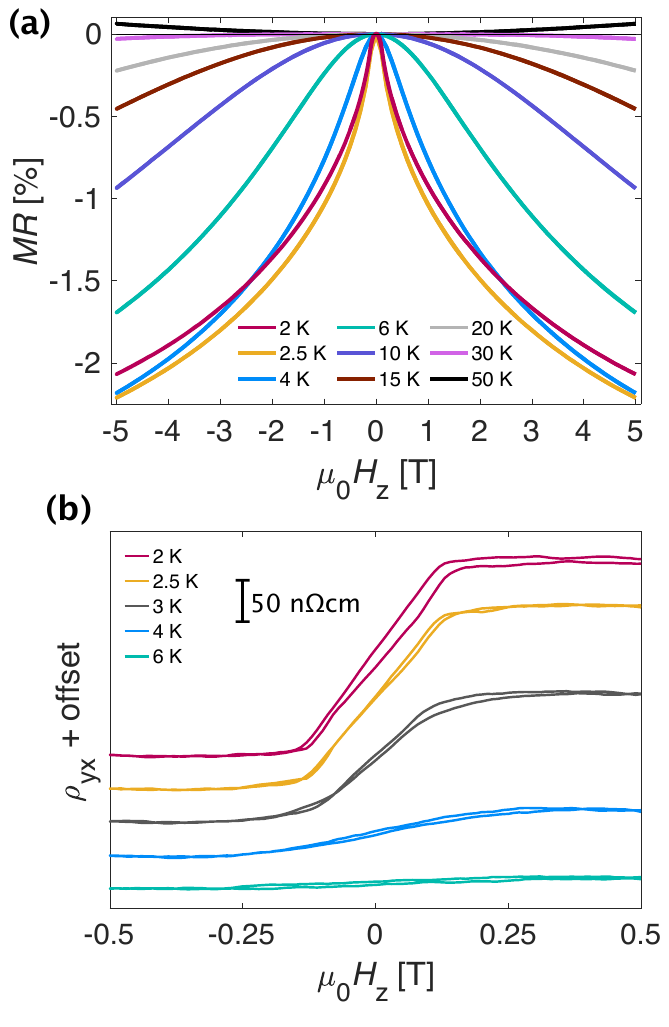}
    \caption{(a) Magnetoresistance ratio versus external magnetic field at several temperatures. (b) Transversal resistivity as a function of magnetic field at some low temperatures. The linear background associated to the ordinary Hall effect is subtracted and an offset is added for better visualisation.}
    \label{fig:Hall}
\end{figure}

 \begin{figure}[b!]
    \centering
    \includegraphics[width=.95\linewidth]{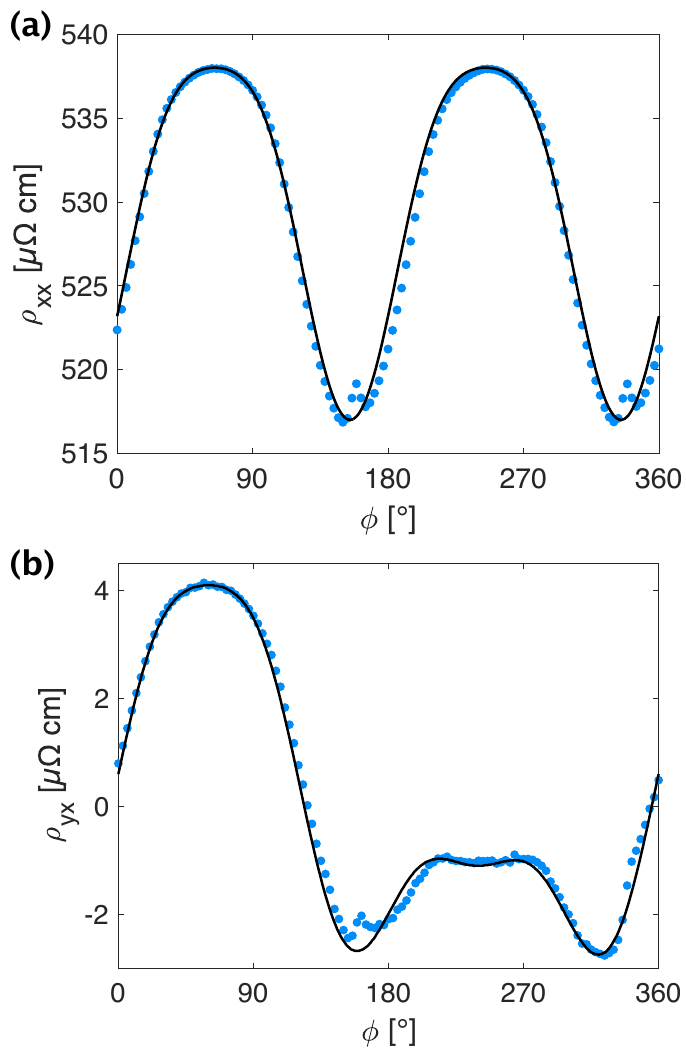}
    \caption{(a) Longitudinal resistivity and (b) transversal resistivity in presence of an in-plane $\SI{5}{\tesla}$-field as a function of the angle between the current line and the magnetic field direction at $\SI{2}{\kelvin}$. The solid black lines show a fitted curve based on the phenomenological relations given in the text for $\rho_{xx}$ and $\rho_{yx}$. The fitted values for the parameters of longitudinal signal are $C_0=521$, $C_2=21$, $C_4=-4$, and $\phi_0=65^\circ$. For the transverse resistivity the fitting coefficients are $C'_1=2.6$, $C'_2=2.0$, $C'_4=0.5$, $\phi'_0=14.5^\circ$, and $\theta_{tilt} = 30^\circ$. 
    The unit of all coefficients $C_i$, and $C'_i$ is $\mu \Omega\,$cm, the same as the resistivities.}  
    \label{fig:rotations}
\end{figure}

To quantify the MR ratio, defined as $(R(H)-R(0))/R(0)$, and the Hall effect, a small dc current of $I = \SI{50}{\micro\ampere}$ along $x$ direction is injected, and the longitudinal ($V_{xx}$) and transversal ($V_{yx}$) voltages are simultaneously measured in an out-of-plane magnetic field sweep. The longitudinal resistivity corresponds to the diagonal components of the resistivity tensor and their variation in magnetic field (magnetoresistivity) is typically an even function of the magnetic field. The transversal signal corresponds to the off-diagonal components of the resistivity tensor which are typically odd under magnetic field reversal. In our measured signals we have both contributions from longitudinal and transversal voltage due to the small misalignment between the electrical contacts on the \ch{PrRhC2} lamella [Fig.~\hyperref[fig:device]{3(a)}]. When an electrical current is applied along high symmetry direction we can isolate the longitudinal and transversal signal separating the measured signal into symmetric and antisymmetric components, respectively. The MR traces measured in an out-of-plane magnetic field at different base temperatures are displayed in Fig.~\hyperref[fig:Hall]{4(a)}. The negative MR at low temperature suggests a weak localization effect, a phenomenon that originates from the constructive interference of backscattered electronic wave functions, that increases the probability to localize the electron \cite{bergmann1984weak}. The coherent superposition of the wave functions is destroyed by applying a magnetic field, resulting in a decrease of the resistivity and thus a negative MR. This effect disappears in our sample above $\SI{30}{\kelvin}$, due to the dominant incoherent scattering at higher temperatures. The positive MR with the characteristic quadratic dependency on the external field above $\SI{30}{\kelvin}$ can be explained by the Lorentz force acting on the carrier motion. The absolute transversal resistivity ($\rho_{yx}$) curves versus the external magnetic field, taken at very low temperatures, are shown in Fig.~\hyperref[fig:Hall]{4(b)}, after subtracting the linear ordinary Hall background with a positive slope of $\SI{1.06}{\micro\ohm\centi\meter\per\tesla}$ . Below $\SI{6}{\kelvin}$, an anomalous contribution to the Hall signal is visible around zero magnetic field, consistent with the temperature dependence of electrical resistivity data [Fig.~\hyperref[fig:device]{3(c)}]. \rev{Assuming a single band model, from the linear region of the $\rho_{yx}$ at high magnetic fields of Fig.~S3 in the Supplementary Information, we estimate the carrier concentration $n_0 = \SI{1.32e21}{\centi\meter^{-3}}$ (electron-type) and the carrier mobility $\mu = \SI{5.7}{\centi\meter^2\per\volt\per\second}$ at $\SI{300}{\kelvin}$.} 

In order to study the in-plane anisotropy, we perform \rev{azimuthal} angular dependent magnetoresistance measurements at $\SI{2}{\kelvin}$ \rev{under an in-plane }magnetic field \rev{with constant} amplitude of $\SI{5}{\tesla}$. The \rev{azimuthal} angular variation of longitudinal resistivity shown in Fig.~\hyperref[fig:rotations]{5(a)} can be well approximated with $\rho_{xx}=C_0+C_2\cos^2 (\phi-\phi_0)+C_4\cos^4 (\phi-\phi_0)$ \cite{Annadi2013}. This equation is a modified version of a well-known phenomenological model for anisotropic magnetoresistance in magnetic systems. The modified equation includes a misalignment angle $\phi_0$ and a fourfold term, both of which have been previously introduced in studying anisotropic magnetoresistance (AMR) in other systems arising from the crystal symmetry \cite{ritzinger2021anisotropic} or heterointerfaces \cite{Awschalom2003,Bason2009,Annadi2013}. The misalignment angle $\phi_0$ can be attributed to the misalignment between magnetization and the external field, as well as the orientation of the crystallographic-axis with respect to the transport measurement geometry. \rev{Typically, in the absence of crystalline anisotropy, rotating the magnetic field within the 2D plane leads to a conventional two-fold symmetric form for the in-plane anisotropic magnetoresistance as a function of the angle between the magnetic field and the current' directions. Our magnetotransport measurement on a single crystal reveals four-fold symmetric contributions, which are consistent and also reflect the underlying crystalline symmetry of the material.}\cite{Annadi2013,Awschalom2008,wadehra2020planar}. Especially at intermediate field limit, a very intricate dependence of the MR as function of the angles between magnetization and current with respect to a reference crystallographic axis has been reported even for simple cubic lattices \cite{Goldstein2017, Takagi2021}. In the present case, the minimal model given above sufficiently mimics the observed AMR, as illustrated by the experimental data. Turning to the transversal signal, we observe transversal AMR, sometimes called planar Hall effect (PHE), together with a cosine-like function as shown in Fig.~\hyperref[fig:rotations]{5(b)}. The PHE was originally proposed and measured in ferromagnetic thin films, where the Hall voltage scales quadratically with the in-plane magnetization as $\rho_{yx} \propto M_{\parallel}^2 \sin 2 \phi_{M,I}$, with $\phi_{M,I}$ representing the angle between the in-plane magnetization and the applied current \cite{McGuire1975,Awschalom2003,Tanaka2008}. Recently, a very strong PHE has been observed in Weyl semimetals due to the chiral anomaly in these systems \cite{Burkov2017,Tewari2017}. In this case, the Hall conductivity depends quadratically on the in-plane magnetic field, and its angular dependence is determined by the angle between the current and applied in-plane magnetic field. Hence, a strong PHE may be observed (i) in FM films with in-plane magnetization and (ii) in Weyl semimetals caused by the chiral anomaly. PrRhC$_2$ is a promising magnetic Weyl semimetal and, due to its canted AFM order, has a FM component with in-plane magnetization. Another complication arises from the role of magnetization in tailoring the Weyl cones \cite{ray2022tunable} which may cause linear terms in the PHE proportional to $B\sin \big(\phi-\theta_{\rm tilt}\big)$ as predicted in Ref.\,\cite{Xie2019}. This quantity could also be given by the ordinary Hall effect resulting from a finite out-of-plane component of the magnetic field and/or due to the unknown crystalline orientation, that might produce any term of the angular dependent transverse resistivity. Taking all the effects into account, we can consider a phenomenological form for the transverse resistivity, $\rho_{yx}=C'_1 \sin \big(\phi+\theta_{\rm tilt}\big) + C'_2 \sin \big[2(\phi-\phi'_0)\big]
+ C'_4 \cos \big[4(\phi-\phi'_0)\big]$, that provides a good agreement with the measured resistivity, as illustrated in Fig.~\hyperref[fig:rotations]{5(b)}. \rev{The angles $\theta_{\rm tilt}$ and $\phi_0'$ account for asymmetries resulting from misalignments between the current direction and the crystallographic orientation of the material. More specifically, $\theta_{\rm tilt}$ can be interpreted as the direction of Weyl nodes tilting relative to the device orientation, while $\phi_0'$ represents the angle between the current direction and the principal in-plane crystal axis.} For the quantified $C'_1= \SI{2.6}{\ohm\centi\meter}$, an out-of-plane misalignment of $\SI{30}{\degree}$ would be necessary for the ordinary Hall effect to be the origin of the linear term, and therefore this scenario can safely be discarded. Excluding the relative reduced AMR ($C'_2 = \SI{2.0}{\ohm\centi\meter} <C'_1$) and a misalignment of the sample, we assume the contribution to the transverse resistivity to be caused by the titled Weyl cones. We emphasize that we are here dealing with a phenomenological model. Indeed, a more elaborate theoretical analysis is needed  to understand the origin of different terms and their connection to the topological Weyl phase of the system along with its magnetic properties.


\vspace{.5em}

\section{Conclusions}
This study investigates the magnetic order of a polycrystal along with thermal and magnetotransport properties in \ch{PrRhC2} single crystals. Our results strongly suggest a non-collinear antiferromagnetic state with a robust net magnetization at low temperature. This ferromagnetic-like behavior is also confirmed by our magnetotransport study, which shows a resistivity peak around $\SI{3}{\kelvin}$ and a significant anomalous Hall signal. We also observe \rev{azimuthal} angular variations in magnetoresistance, indicating the presence of higher-order terms in the resistance tensor and potentially Weyl phase character of the material. Our findings highlight the potential of \ch{PrRhC2}, as a unique material to study the interplay between magnetic and topological semimetallic properties, particularly due to the breaking of both time-reversal and inversion symmetries.

\section{Acknowledgments}
We are indebted to Dr. Rajyavardhan Ray and Dr. Jorge Facio for fruitful discussions. We thank the Collaborative Research Center SFB 1143 (Project No. A05), and the W\"urzburg-Dresden Cluster of Excellence on Complexity and Topology in Quantum Matter – ct.qmat (EXC 2147, Project No. 390858490) for support. H.R. acknowledges GACR 22-17899K. DK acknowledges the Lumina Quaeruntur fellowship LQ100102201 of the Czech Academy of Sciences and grant number 22-22000M from the Czech Science Foundation. Both L.T.C. and V.R. are funded by the DFG (project 456950766 (L.T.C.) and project RO 6386/2-1 (V.R.)).


\begin{thebibliography}{}

\bibitem{wan2011topological}Wan, X., Turner, A., Vishwanath, A. \& Savrasov, S. Topological semimetal and Fermi-arc surface states in the electronic structure of pyrochlore iridates. {\em Phys.~Rev.~B}. \textbf{83}, 205101 (2011)
\bibitem{burkov2016topological}Burkov, A. Topological semimetals. {\em Nat.~Mater.}. \textbf{15}, 1145-1148 (2016)
\bibitem{liu2014weyl}Liu, J. \& Vanderbilt, D. Weyl semimetals from noncentrosymmetric topological insulators. {\em Phys.~Rev.~B}. \textbf{90}, 155316 (2014)
\bibitem{sie2019ultrafast}Sie, E., Nyby, C., Pemmaraju, C., Park, S., Shen, X., Yang, J., Hoffmann, M., Ofori-Okai, B., Li, R., Reid, A. \& Others An ultrafast symmetry switch in a Weyl semimetal. {\em Nature}. \textbf{565}, 61-66 (2019)
\bibitem{burkov2011weyl}Burkov, A. \& Balents, L. Weyl semimetal in a topological insulator multilayer. {\em Phys.~Rev.~Lett.}. \textbf{107}, 127205 (2011)
\bibitem{wang2012dirac}Wang, Z., Sun, Y., Chen, X., Franchini, C., Xu, G., Weng, H., Dai, X. \& Fang, Z. Dirac semimetal and topological phase transitions in A\textsubscript{3}Bi (A= Na, K, Rb). {\em Phys.~Rev.~B}. \textbf{85}, 195320 (2012)
\bibitem{weng2015weyl}Weng, H., Fang, C., Fang, Z., Bernevig, B. \& Dai, X. Weyl semimetal phase in noncentrosymmetric transition-metal monophosphides. {\em Phys.~Rev.~X}. \textbf{5}, 011029 (2015)
\bibitem{yan2017topological}Yan, B. \& Felser, C. Topological materials: Weyl semimetals. {\em Annu.~Rev.~Condens.~Matter Phys.}. \textbf{8} pp. 337-354 (2017)
\bibitem{armitage2018weyl}Armitage, N., Mele, E. \& Vishwanath, A. Weyl and Dirac semimetals in three-dimensional solids. {\em Rev.~Mod.~Phys.}. \textbf{90}, 015001 (2018)
\bibitem{morali2019fermi}Morali, N., Batabyal, R., Nag, P., Liu, E., Xu, Q., Sun, Y., Yan, B., Felser, C., Avraham, N. \& Beidenkopf, H. Fermi-arc diversity on surface terminations of the magnetic Weyl semimetal Co\textsubscript{3}Sn\textsubscript{2}S\textsubscript{2}. {\em Science}. \textbf{365}, 1286-1291 (2019)
\bibitem{liu2019magnetic}Liu, D., Liang, A., Liu, E., Xu, Q., Li, Y., Chen, C., Pei, D., Shi, W., Mo, S., Dudin, P. \& Others Magnetic Weyl semimetal phase in a Kagomé crystal. {\em Science}. \textbf{365}, 1282-1285 (2019)
\bibitem{shen2019anisotropies}Shen, J., Zeng, Q., Zhang, S., Tong, W., Ling, L., Xi, C., Wang, Z., Liu, E., Wang, W., Wu, G. \& Others On the anisotropies of magnetization and electronic transport of magnetic Weyl semimetal Co\textsubscript{3}Sn\textsubscript{2}S\textsubscript{2}. {\em Appl.~Phys.~Lett.}. \textbf{115}, 212403 (2019)
\bibitem{okamura2020giant}Okamura, Y., Minami, S., Kato, Y., Fujishiro, Y., Kaneko, Y., Ikeda, J., Muramoto, J., Kaneko, R., Ueda, K., Kocsis, V. \& Others Giant magneto-optical responses in magnetic Weyl semimetal Co3Sn2S2. {\em Nat.~Comm.}. \textbf{11}, 1-8 (2020)
\bibitem{nie2022magnetic}Nie, S., Hashimoto, T. \& Prinz, F. Magnetic Weyl Semimetal in K\textsubscript{2}Mn\textsubscript{3}(AsO\textsubscript{4})\textsubscript{3} with the Minimum Number of Weyl Points. {\em Phys.~Rev.~Lett.}. \textbf{128}, 176401 (2022)

\bibitem{su2020magnetic}Su, H., Gong, B., Shi, W., Yang, H., Wang, H., Xia, W., Yu, Z., Guo, P., Wang, J., Ding, L. \& Others Magnetic exchange induced Weyl state in a semimetal EuCd\textsubscript{2}Sb\textsubscript{2}. {\em APL Mater.}. \textbf{8}, 011109 (2020)
\bibitem{liu2018giant}Liu, E., Sun, Y., Kumar, N., Muechler, L., Sun, A., Jiao, L., Yang, S., Liu, D., Liang, A., Xu, Q. \& Others Giant anomalous Hall effect in a ferromagnetic kagome-lattice semimetal. {\em Nat.~Phys.}. \textbf{14}, 1125-1131 (2018)
\bibitem{kuroda2017evidence}Kuroda, K., Tomita, T., Suzuki, M., Bareille, C., Nugroho, A., Goswami, P., Ochi, M., Ikhlas, M., Nakayama, M., Akebi, S. \& Others Evidence for magnetic Weyl fermions in a correlated metal. {\em Nat.~Mater.}. \textbf{16}, 1090-1095 (2017)
\bibitem{kim2018large}Kim, K., Seo, J., Lee, E., Ko, K., Kim, B., Jang, B., Ok, J., Lee, J., Jo, Y., Kang, W. \& Others Large anomalous Hall current induced by topological nodal lines in a ferromagnetic van der Waals semimetal. {\em Nat.~Mater.}. \textbf{17}, 794-799 (2018)
\bibitem{ray2022tunable}Ray, R., Sadhukhan, B., Richter, M., Facio, J. \& Brink, J. Tunable chirality of noncentrosymmetric magnetic Weyl semimetals in rare-earth carbides. {\em Npj Quantum Mater.}. \textbf{7}, 1-9 (2022)
\bibitem{sadhukhan2023effect}Sadhukhan, B. \& Nag, T. Effect of chirality imbalance on Hall transport of PrRhC\textsubscript{2}. {\em Phys.~Rev.~B}. \textbf{107}, L081110 (2023)

\bibitem{zhang2022weyl}Zhang, R., Huang, C., Kidd, J., Markiewicz, R., Lin, H., Bansil, A., Singh, B. \& Sun, J. Weyl semimetal in the rare-earth hexaboride family supporting a pseudonodal surface and a giant anomalous Hall effect. {\em Phys.~Rev.~B}. \textbf{105}, 165140 (2022)
\bibitem{yang2021noncollinear}Yang, H., Singh, B., Gaudet, J., Lu, B., Huang, C., Chiu, W., Huang, S., Wang, B., Bahrami, F., Xu, B. \& Others Noncollinear ferromagnetic Weyl semimetal with anisotropic anomalous Hall effect. {\em Phys.~Rev.~B}. \textbf{103}, 115143 (2021)
\bibitem{yang2020transition}Yang, H., Singh, B., Lu, B., Huang, C., Bahrami, F., Chiu, W., Graf, D., Huang, S., Wang, B., Lin, H. \& Others Transition from intrinsic to extrinsic anomalous Hall effect in the ferromagnetic Weyl semimetal PrAlGe\textsubscript{1-$x$}Si\textsubscript{$x$}. {\em Apl Materials}. \textbf{8}, 011111 (2020)
\bibitem{chang2018magnetic}Chang, G., Singh, B., Xu, S., Bian, G., Huang, S., Hsu, C., Belopolski, I., Alidoust, N., Sanchez, D., Zheng, H. \& Others Magnetic and noncentrosymmetric Weyl fermion semimetals in the R AlGe family of compounds (R= rare earth). {\em Phys.~Rev.~B}. \textbf{97}, 041104 (2018)
\bibitem{xu2017discovery}Xu, S., Alidoust, N., Chang, G., Lu, H., Singh, B., Belopolski, I., Sanchez, D., Zhang, X., Bian, G., Zheng, H. \& Others Discovery of Lorentz-violating type II Weyl fermions in LaAlGe. {\em Sci.~Adv.}. \textbf{3}, e1603266 (2017)
\bibitem{nielsen1983adler}Nielsen, H. \& Ninomiya, M. The Adler-Bell-Jackiw anomaly and Weyl fermions in a crystal. {\em Phys.~Lett.~B}. \textbf{130}, 389-396 (1983)
\bibitem{zyuzin2012topological}Zyuzin, A. \& Burkov, A. Topological response in Weyl semimetals and the chiral anomaly. {\em Phys.~Rev.~B}. \textbf{86}, 115133 (2012)
\bibitem{son2013chiral}Son, D. \& Spivak, B. Chiral anomaly and classical negative magnetoresistance of Weyl metals. {\em Phys.~Rev.~B}. \textbf{88}, 104412 (2013)
\bibitem{huang2015observation}Huang, X., Zhao, L., Long, Y., Wang, P., Chen, D., Yang, Z., Liang, H., Xue, M., Weng, H., Fang, Z. \& Others Observation of the chiral-anomaly-induced negative magnetoresistance in 3D Weyl semimetal TaAs. {\em Phys.~Rev.~X}. \textbf{5}, 031023 (2015)
\bibitem{jia2016weyl}Jia, S., Xu, S. \& Hasan, M. Weyl semimetals, Fermi arcs and chiral anomalies. {\em Nature Materials}. \textbf{15}, 1140-1144 (2016)
\bibitem{zhang2017tunable}Zhang, E., Chen, R., Huang, C., Yu, J., Zhang, K., Wang, W., Liu, S., Ling, J., Wan, X., Lu, H. \& Others Tunable positive to negative magnetoresistance in atomically thin WTe\textsubscript{2}. {\em Nano Lett.}. \textbf{17}, 878-885 (2017)
\bibitem{goswami2015axial}Goswami, P., Pixley, J. \& Sarma, S. Axial anomaly and longitudinal magnetoresistance of a generic three-dimensional metal. {\em Phys.~Rev.~B}. \textbf{92}, 075205 (2015)
\bibitem{schumann2017negative}Schumann, T., Goyal, M., Kealhofer, D. \& Stemmer, S. Negative magnetoresistance due to conductivity fluctuations in films of the topological semimetal Cd\textsubscript{3}As\textsubscript{2}. {\em Phys.~Rev.~B}. \textbf{95}, 241113 (2017)
\bibitem{dos2016search}Dos Reis, R., Ajeesh, M., Kumar, N., Arnold, F., Shekhar, C., Naumann, M., Schmidt, M., Nicklas, M. \& Hassinger, E. On the search for the chiral anomaly in Weyl semimetals: the negative longitudinal magnetoresistance. {\em New J.~Phys.}. \textbf{18}, 085006 (2016)
\bibitem{arnold2016negative}Arnold, F., Shekhar, C., Wu, S., Sun, Y., Dos Reis, R., Kumar, N., Naumann, M., Ajeesh, M., Schmidt, M., Grushin, A. \& Others Negative magnetoresistance without well-defined chirality in the Weyl semimetal TaP. {\em Nat.~Comm.}. \textbf{7}, 1-7 (2016)
\bibitem{matsuo1996antiferromagnetism}Matsuo, S., Onodera, H., Kosaka, M., Kobayashi, H., Ohashi, M., Yamauchi, H. \& Yamaguchi, Y. Antiferromagnetism of GdCoC\textsubscript{2} and GdNiC\textsubscript{2} intermetallics studied by magnetization measurement and 155Gd Mössbauer spectroscopy. {\em J.~Magn.~Magn.~Mater.}. \textbf{161} pp. 255-264 (1996)
\bibitem{meng2016magnetic}Meng, L., Xu, C., Yuan, Y., Qi, Y., Zhou, S. \& Li, L. Magnetic properties and giant reversible magnetocaloric effect in GdCoC\textsubscript{2}. {\em Rsc Advances}. \textbf{6}, 74765-74768 (2016)
\bibitem{onodera1998magnetic}Onodera, H., Koshikawa, Y., Kosaka, M., Ohashi, M., Yamauchi, H. \& Yamaguchi, Y. Magnetic properties of single-crystalline RNiC\textsubscript{2} compounds (R= Ce, Pr, Nd and Sm). {\em J.~Magn.~Magn.~Mater.}. \textbf{182}, 161-171 (1998)
\bibitem{hoffmann1989structural}Hoffmann, R., Jeitschko, W. \& Boonk, L. Structural, chemical, and physical properties of rare-earth metal rhodium carbides LnRhC\textsubscript{2} (Ln= La, Ce, Pr, Nd, Sm). {\em Chem.~Mater.}. \textbf{1}, 580-586 (1989)
\bibitem{chen2014anomalous}Chen, H., Niu, Q. \& MacDonald, A. Anomalous Hall effect arising from noncollinear antiferromagnetism. {\em Phys.~Rev.~Lett.}. \textbf{112}, 017205 (2014)
\bibitem{nakatsuji2015large}Nakatsuji, S., Kiyohara, N. \& Higo, T. Large anomalous Hall effect in a non-collinear antiferromagnet at room temperature. {\em Nature}. \textbf{527}, 212-215 (2015)
\bibitem{Blundell}{S.Blundell, Magnetism in Condensed Matter (Oxford University Press, 2001)}
\bibitem{geishendorf2019magnetoresistance}Geishendorf, K., Schlitz, R., Vir, P., Shekhar, C., Felser, C., Nielsch, K., Goennenwein, S. \& Thomas, A. Magnetoresistance and anomalous Hall effect in micro-ribbons of the magnetic Weyl semimetal Co3Sn2S2. {\em Applied Physics Letters}. \textbf{114} (2019)
\bibitem{kataoka2001resistivity}Kataoka, M. Resistivity and magnetoresistance of ferromagnetic metals with localized spins. {\em Phys.~Rev.~B}. \textbf{63}, 134435 (2001)
\bibitem{bergmann1984weak}Bergmann, G. Weak localization in thin films: a time-of-flight experiment with conduction electrons. {\em Phys.~Rep.}. \textbf{107}, 1-58 (1984)
\bibitem{Annadi2013}Annadi, A., Huang, Z., Gopinadhan, K., Wang, X., Srivastava, A., Liu, Z., Ma, H., Sarkar, T., Venkatesan, T. \& Ariando Fourfold oscillation in anisotropic magnetoresistance and planar Hall effect at the LaAlO\textsubscript{3}/SrTiO\textsubscript{3} heterointerfaces: Effect of carrier confinement and electric field on magnetic interactions. {\em Phys.~Rev.~B}. \textbf{87}, 201102 (2013,5)
\bibitem{ritzinger2021anisotropic}Ritzinger, P., Reichlova, H., Kriegner, D., Markou, A., Schlitz, R., Lammel, M., Scheffler, D., Park, G., Thomas, A., Středa, P. \& Others Anisotropic magnetothermal transport in Co\textsubscript{2}MnGa thin films. {\em Phys.~Rev.~B}. \textbf{104}, 094406 (2021)
\bibitem{Awschalom2003}Tang, H., Kawakami, R., Awschalom, D. \& Roukes, M. Giant Planar Hall Effect in Epitaxial (Ga,Mn)As Devices. {\em Phys.~Rev.~Lett.}. \textbf{90}, 107201 (2003,3)
\bibitem{Bason2009}Bason, Y., Hoffman, J., Ahn, C. \& Klein, L. Magnetoresistance tensor of La\textsubscript{0.8}Sr\textsubscript{0.2}MnO\textsubscript{3}. {\em Phys. Rev. B}. \textbf{79}, 092406 (2009,3)
\bibitem{Awschalom2008}Wu, D., Wei, P., Johnston-Halperin, E., Awschalom, D. \& Shi, J. High-field magnetocrystalline anisotropic resistance effect in (Ga,Mn)As. {\em Phys.~Rev.~B}. \textbf{77}, 125320 (2008,3)
\bibitem{wadehra2020planar}Wadehra, N., Tomar, R., Varma, R., Gopal, R., Singh, Y., Dattagupta, S. \& Chakraverty, S. Planar Hall effect and anisotropic magnetoresistance in polar-polar interface of LaVO\textsubscript{3}-KTaO\textsubscript{3} with strong spin-orbit coupling. {\em Nat.~Commun.}. \textbf{11}, 874 (2020)
\bibitem{Goldstein2017}Rout, P., Agireen, I., Maniv, E., Goldstein, M. \& Dagan, Y. Six-fold crystalline anisotropic magnetoresistance in the (111) LaAlO\textsubscript{3}/SrTiO\textsubscript{3} oxide interface. {\em Phys.~Rev.~B}. \textbf{95}, 241107 (2017,6)
\bibitem{Takagi2021}Huang, D., Nakamura, H. \& Takagi, H. Planar Hall effect with sixfold oscillations in a Dirac antiperovskite. {\em Phys. Rev. Res.}. \textbf{3}, 013268 (2021,3)
\bibitem{McGuire1975}McGuire, T. \& Potter, R. Anisotropic magnetoresistance in ferromagnetic 3d alloys. {\em EEE Trans.~Magn.}. \textbf{11}, 1018 (1975)
\bibitem{Tanaka2008}Nazmul, A., Lin, H., Tran, S., Ohya, S. \& Tanaka, M. Planar Hall effect and uniaxial in-plane magnetic anisotropy in Mn $\delta$ -doped GaAs $p$-AlGaAs heterostructures. {\em Phys. Rev. B}. \textbf{77}, 155203 (2008,4)
\bibitem{Burkov2017}Burkov, A. Giant planar Hall effect in topological metals. {\em Phys.~Rev.~B}. \textbf{96}, 041110 (2017,7)
\bibitem{Tewari2017}Nandy, S., Sharma, G., Taraphder, A. \& Tewari, S. Chiral Anomaly as the Origin of the Planar Hall Effect in Weyl Semimetals. {\em Phys. Rev. Lett.}. \textbf{119}, 176804 (2017,10)
\bibitem{Xie2019}Ma, D., Jiang, H., Liu, H. \& Xie, X. Planar Hall effect in tilted Weyl semimetals. {\em Phys.~Rev.~B}. \textbf{99}, 115121 (2019,3)




\end{thebibliography}
\bibliographystyle{}

\end{document}